\documentclass[authoryear,preprint,review,12pt]{elsarticle}

\usepackage{graphics}
\usepackage{graphicx}
\usepackage{epsfig}
\usepackage{caption}
\usepackage{amssymb}
\usepackage{amsthm}
\usepackage{color}

\def\astrobj#1{#1}
\journal{New Astronomy}

\begin{document}

\begin{frontmatter}


\title{Photometric Solutions of Some Contact ASAS Binaries}

\author{\.{I}.Gezer\corref{dip}$^{1}$}
\author{Z. Bozkurt$^{1}$}
\address{$^{1}$Astronomy and Space Science Department, Ege University, 35100 Bornova, Izmir, Turkey}
\cortext[dip]{Corresponding author \\
E-mail address: gezer.ilknur@gmail.com}

\begin{abstract}
We present the first light curve solution of 6 contact binary systems which are chosen from the ASAS catalog. The photometric elements and the estimated absolute parameters of all systems are obtained with the light curve analyses. We calculated the values of degree of contact for the systems. The location of the targets on the Hertzsprung--Russell diagram and the mass--radius plane is compared to the other well-known contact binaries and the evolutionary status of the systems are also discussed. 
\end{abstract}


\end{frontmatter}

\section{Introduction}

ASAS (All Sky Automated Survey, \citep{poj97}) is a project which aims to detect any kind of photometric variability by monitoring the large area of the sky with fully automated instruments. One of the main objectives of ASAS is to find and catalog variable stars. Through the project, approximately 10$^7$ stars which are brighter than 14$^m$ have been observed so far. The prototype of the project was first operated in 1996 at the Warsaw University Astronomical Observatory. Now, it carries on with three full automatic instruments having $V$ and $I$ filters attached to  CCD cameras at Las Campanas Observatory in Chile and at Mt. Haleakala Observatory in Maui, Hawaii. The categorized stars are relatively located in the southern hemisphere ($\delta < +28^{\circ}$) and many of them are newly discovered. The public domain data of the ASAS also ease the achievement and investigation of the systems in detail. ASAS apparent magnitudes were transformed into the standard 'I ' and 'V ' systems using \cite{Lan92} and Hipparcos \cite{Perry97}. The photometric accuracy is given about 0.05 mag in most cases. 

\cite{poj00} published the first results of observations obtained by the prototype ASAS camera and  gave a catalog containing 3800 variable stars. According to \cite{pac06} 11076 eclipsing binaries (including 5384 contact systems) were discovered. They presented the preliminary results of the analysis for thousands of binary systems. They also emphasized that their statistical investigation supports the hypothesis in which the thermal relaxation oscillation states of contact binaries \citep{fla76,luc76}.

The targets in our study were selected from the eclipsing binary list of \cite{pac06}. We chose our targets according to the criteria that no detailed investigation can be found in literature. The main properties of the targets are listed in Table~\ref{tab1}. In the next section we present the temperature determination method for primary components and the details of the light curve analyses. In section 3, we give detailed information about the targets and some crucial parameters obtained by the light curve solutions. We conclude the results and compare the evolutionary status of the targets to known contact binaries in the last section.

\begin{table*}
\begin{center}
\tiny
\caption{Properties of the selected targets. RA, DEC and T$_0$ refer to the right ascension, the declination and the time of minimum light, respectively. V$_{max}$ and V$_{amp}$ denote the maximum brightness and the amplitude of variation in $V$ filter.}
\renewcommand{\arraystretch}{1.2}
\label{tab1}
\begin{tabular}{lccccccccc}
\hline
ASAS Number    &Other ID&RA($^{h}~^{m}~^{s}$)  &  DEC($^{\circ}~^{m}~^{s}$) & T$_0$(HJD-2450000)  &  Period($^d$) & V$_{max}$($^m$) & V$_{amp}$($^m$) \\
\hline
002821-1453.3 & TYC 5268-1013-1 & 00 28 21 & -14 53 18 & 1869.060 & 0.402660 & 11.54 & 0.44   \\  
012450-3241.4 & TYC 7002-320-1 & 01 24 50 &	-32 41 24 & 1869.099 &	0.308971	& 11.45 &	 0.58      \\
024155+2507.8 & TYC 1772-674-1 & 02 41 55 & 25 07 48 & 2621.660 & 0.400889 & 11.73 & 0.61 \\
050334-2521.9 & TYC 6477-224-1 & 05 03 34 &	-25 21 54 & 1868.980 &	0.414060	& 11.09 &	0.31	\\
051353-1701.2 & TYC 5906-87-1 & 05 13 53 & -17 01 12 & 1869.140 & 0.341836 & 11.66 & 0.55 \\
063546+1928.6 & TYC 1337-1137-1 & 06 35 46	& 19 28 36 & 2621.780 & 0.475511 & 9.95 & 0.43 \\
\hline
\end{tabular}
\end{center}
\end{table*}

\begin{table*}
\begin{center}
\caption{The extinction ratios of different photometric systems given by \citep {Ram05} }
\renewcommand{\arraystretch}{1.2}
\label{tab2}
\begin{tabular}{lcc}
\hline
\,\,\,Color&Photometric system& $k^{a}$  \\
\hline
(V -- $J_{2}$) & Johnson-2MASS & 2.16   \\  
(V -- $H_{2}$)  & Johnson-2MASS & 2.51      \\
(V -- $K_{2}$)  & Johnson-2MASS & 2.70  \\
 ($B_{T}$ -- $V_{T}$) & Tycho & 1.02 	\\
($V_{T}$ -- $K_{2}$) & Tycho-2MASS & 2.87 \\
\hline
$^{a}$\it{ k = E(color) / E(B-V)}
\end{tabular}
\end{center}
\end{table*}

\section{Light curve analyses of programme stars}

\subsection{Temperature determination}
Since the light curve analysis of an eclipsing binary system requires the effective temperature of at least one of the components, the accurate determination of the effective temperature is a critical step in the solution. Our program stars do not have any detailed spectroscopic or photometric study in the literature, however, their Johnson, 2MASS and Tycho magnitudes are given in several data archives. In this case the only way to determine their temperature is to use different colors and temperature calibrations. We used the calibrations given by \cite {Ram05} who listed the adopted extinction ratios for various photometric systems. For the reddening correction, we used calibrations given in Table~\ref{tab2} \citep {Ram05} and $\it {E(B - V)}$ values which are obtained from Kurucz models \citep {kurucz03}. The $\it{(V - K)}$ color was decided to use in the light curve solution because of its relatively low dependence on the metallicity. We then determined the effective temperatures of primary components for derived intrinsic colour indices by using the appropriate table (Table~11) of \cite {Ram05}. During our calculations, we assumed that the primary component is a main-sequence star and its metallicity value is equal to solar metallicity, [Fe/H] = +0.0. Finally, the average values of two calculated temperatures corresponding to ($\it{V-K_{2}}$) and ($\it{V_{T}-K_{2}}$) colors of each target adopted as the effective temperature values for primary components.

\subsection{Analyses}
\begin{figure*}
\begin{center}
\begin{tabular}{|c|c|c|}

\hline
\includegraphics[scale=0.3]{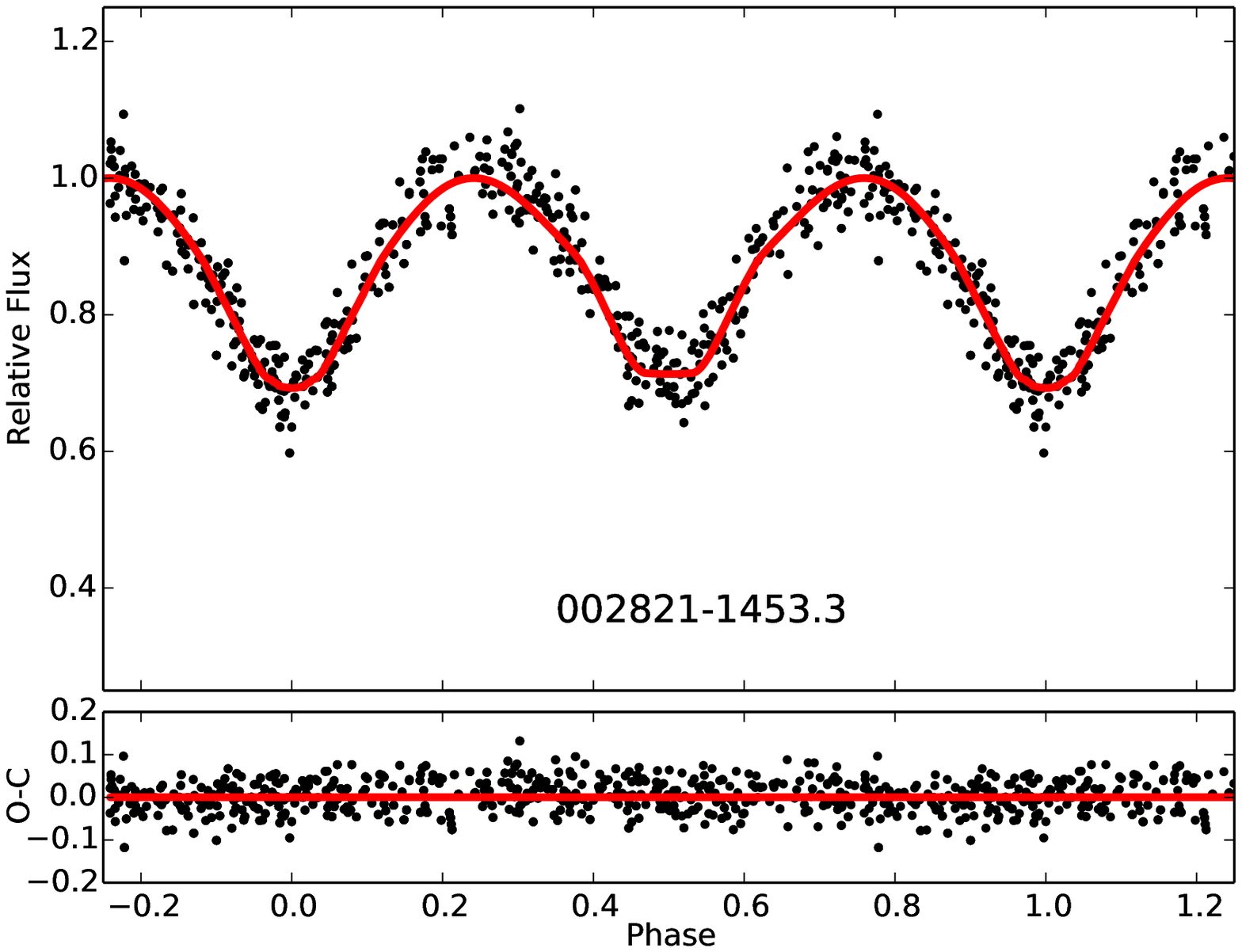} & 
\includegraphics[scale=0.3]{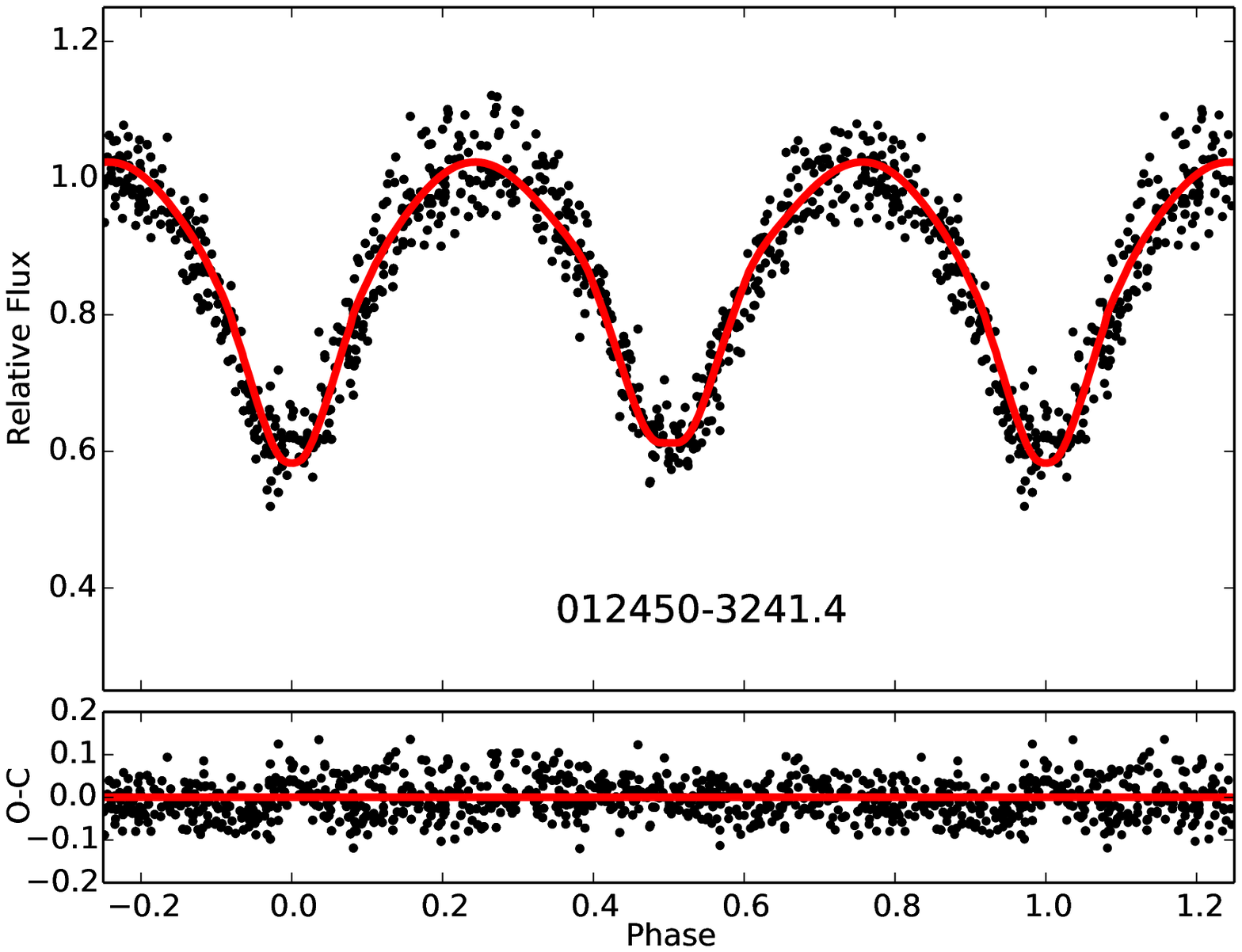} \\
\hline
\includegraphics[scale=0.3]{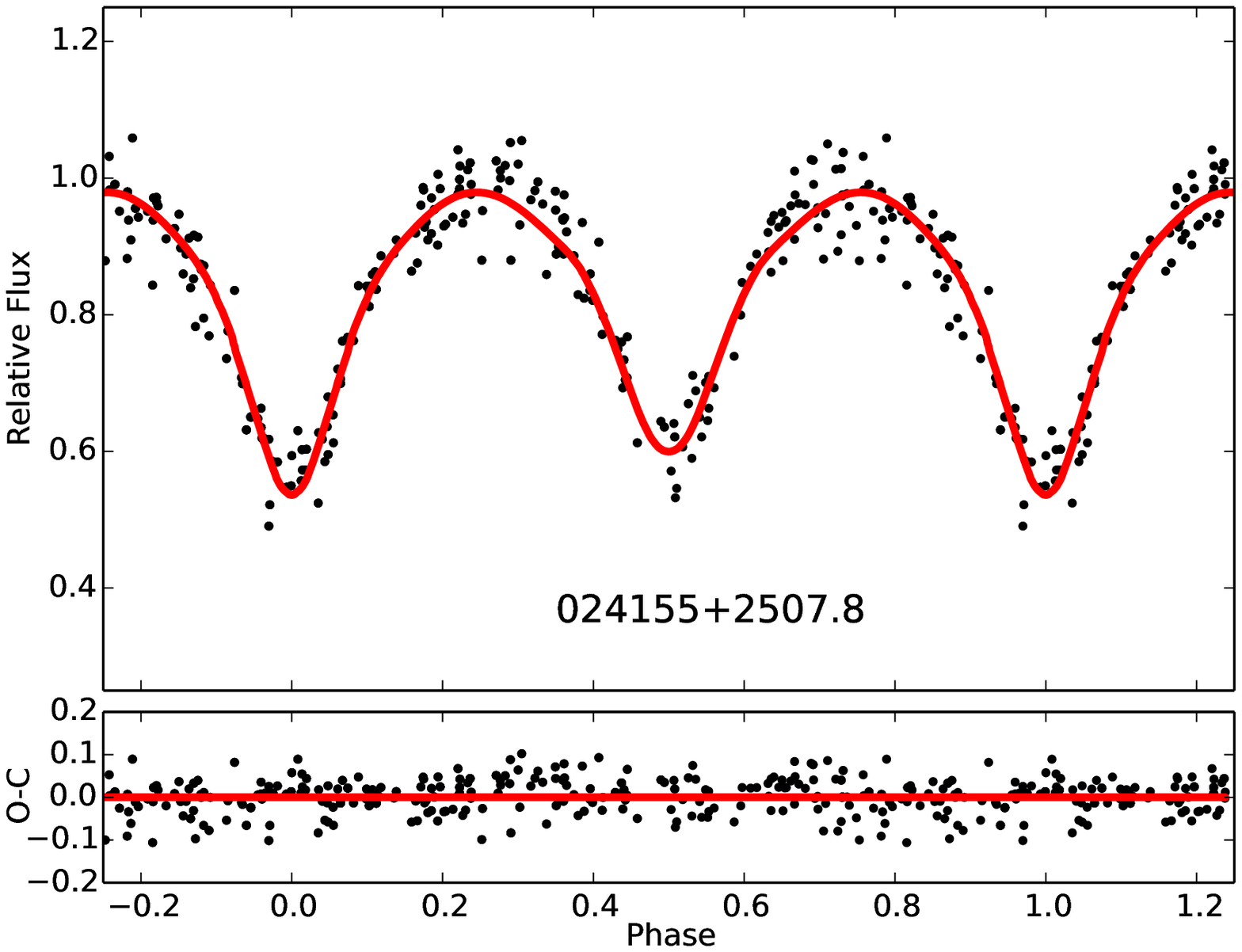}  & 
\includegraphics[scale=0.3]{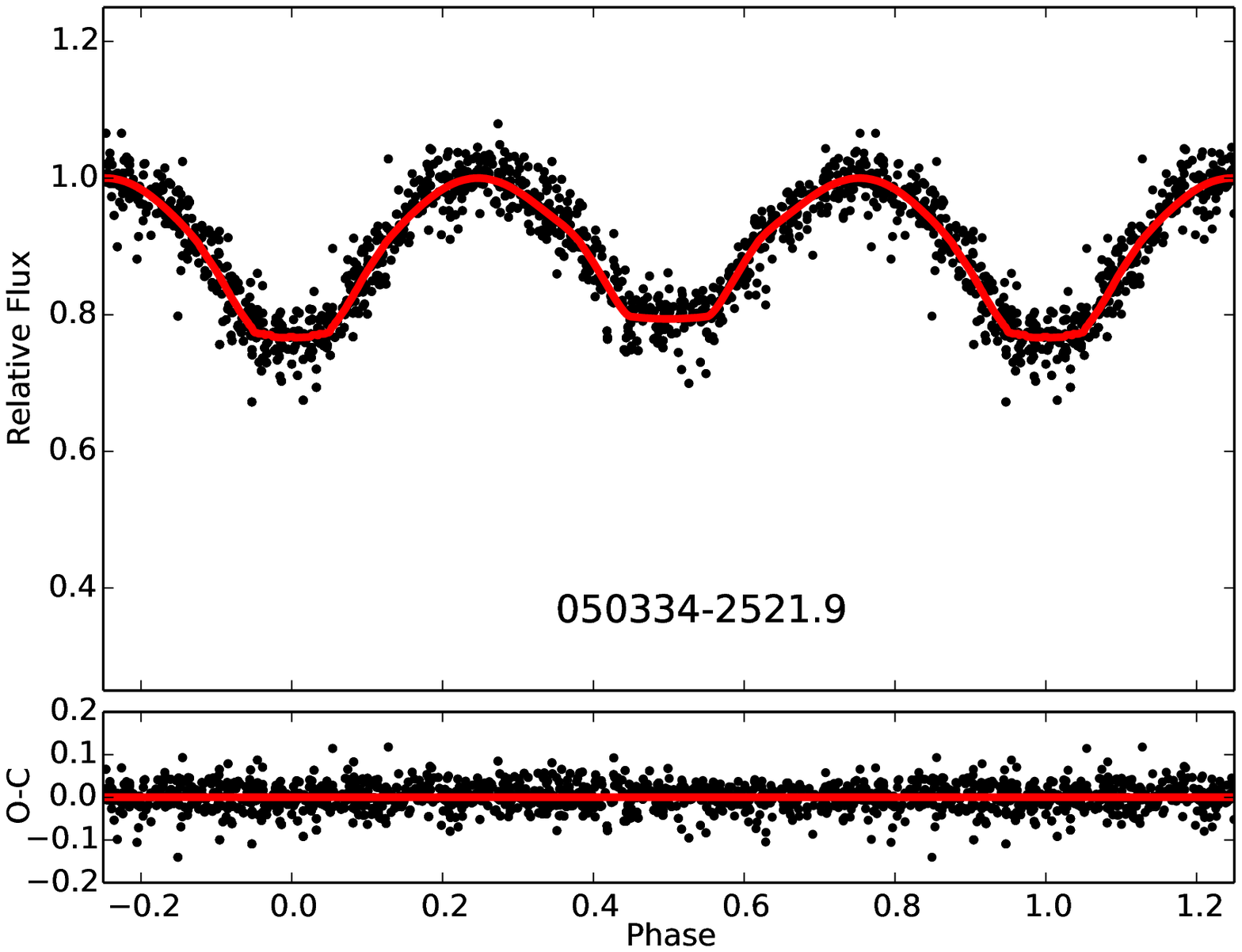} \\
\hline
\includegraphics[scale=0.3]{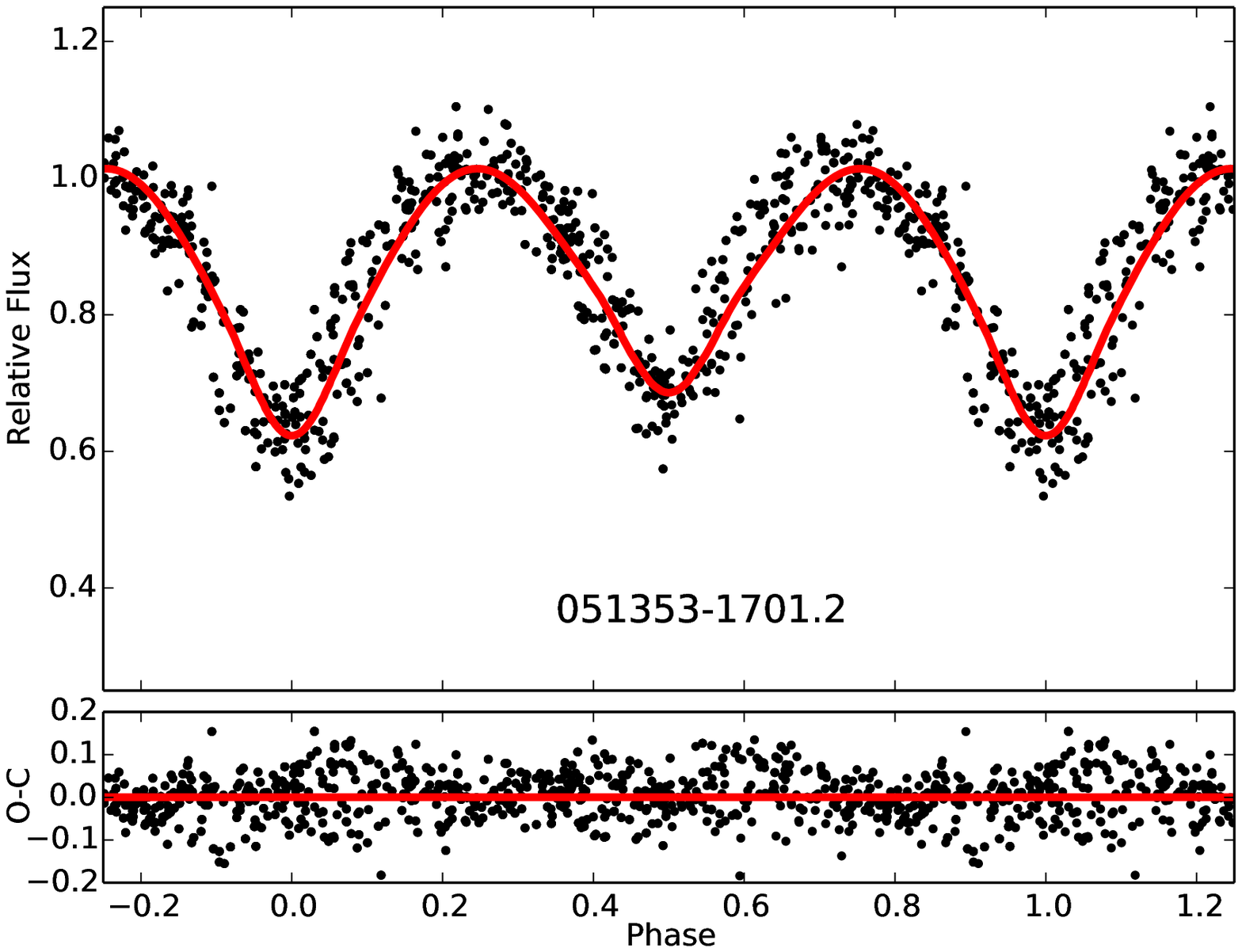} & 
\includegraphics[scale=0.3]{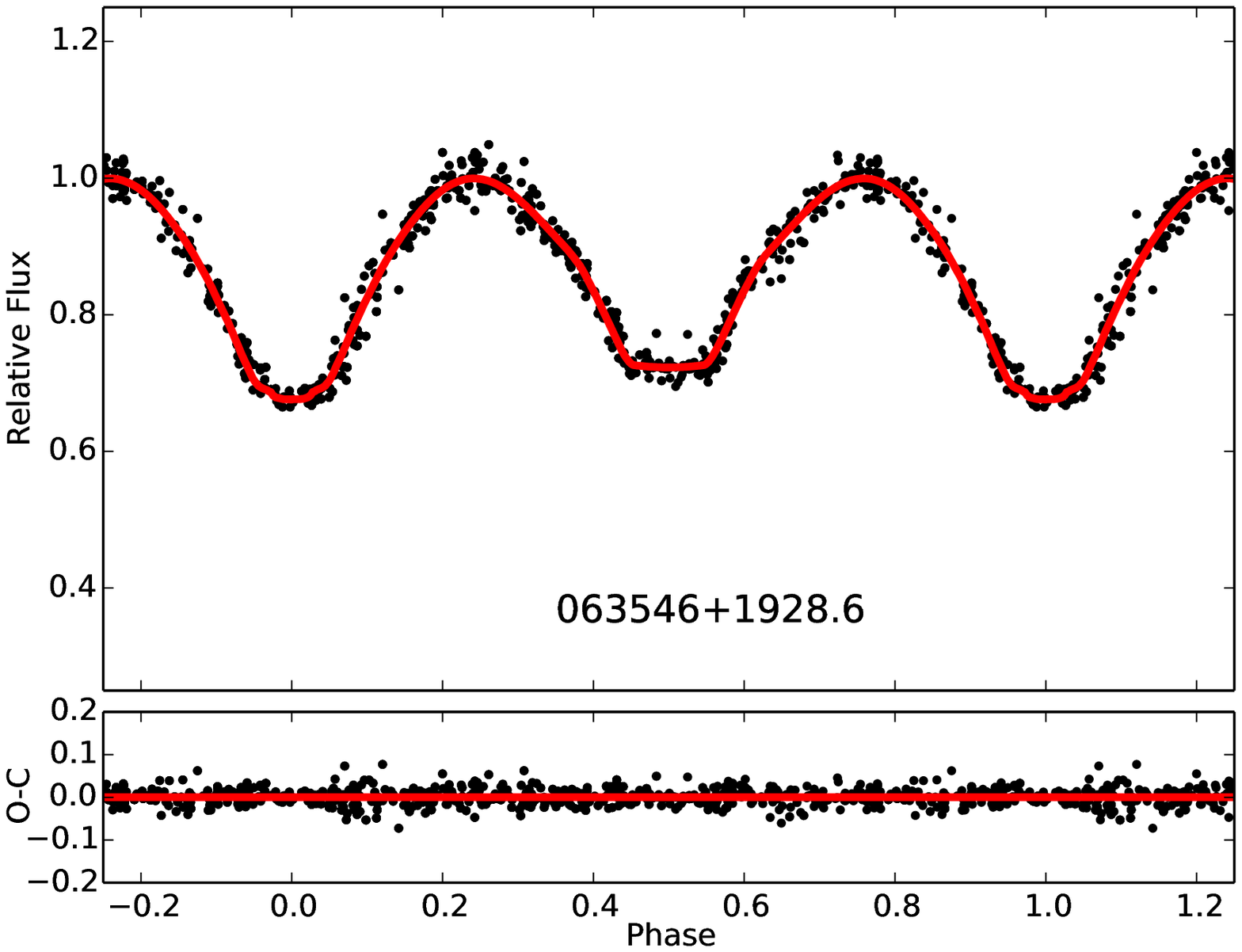} \\
\hline
\end{tabular}
\caption{Agreement between the observational and calculated light curves. Dots indicate observations and lines refer theoretical curves. The ASAS number of the systems are written on the figures. At the bottom of each figure the differences between observational and calculated light curves are also shown.}
\label{figlc}
\end{center}
\end{figure*}

The $V$--band light curves of selected contact binary systems were collected from the ASAS database. We analysed all light curves by using the PHOEBE \cite{prs05} software, which is based on the Wilson--Devinney code \cite{wil71}. The unmeasured and grade D (noted as useless) data were not included to our analyses. The gravitational darkening coefficients of the primary and secondary components, g$_1$ and g$_2$, were chosen from \cite{ruc69} while the albedo values, A$_1$ and A$_2$, were taken from \cite{ham93}. The following parameters were set as adjustable during the light curve analysis: The inclination of the orbit, $i$, mass ratio $q=M_{2}/M_{1}$, temperature of the secondary component, $T_2$, dimensionless surface potentials of the primary and secondary components, $\Omega_1 = \Omega_2$, unnormalized monochromatic luminosity of the primary component,$L_1$, the time of primary minimum, $T_0$, and the orbital period of the binary, $P$. Additionally, we calculated the fillout factor of each system by using the following equation \citep{luc79},

\begin{equation}
f=\frac{\Omega_{i}-\Omega}{\Omega_{i}-\Omega_{o}}, 
\end{equation}
where $\Omega_{i}$ is the inner and $\Omega_{o}$ is the outer Lagrangian potentials. As a first step, we tried to find a solution for all light curves including all observational points (including those with large scattered ones) from the ASAS database. When we reach a reasonable solution, we calculated the differences between the theoretical and observational light curves points for this initial solution. In order to extract the useless scattered points we calculated the standard deviations, $\sigma$, of these differences and we removed the points located beyond the 3$\sigma$. Then we run the code again to find the final solution. The resulting parameters yielded by analysis of 6 systems including output errors of PHOEBE code are presented in Table~\ref{tablc} and the theoretical light curves among the observational points are also shown in Fig.~\ref{figlc}. At the bottom of the figures we also present the differences between computed and observational light curves in the final solution. In the following subsections, we give the some details of the systems based on our analyses.

\begin{table*}
\begin{center}
\caption{Results of the light curve analysis of the systems. Formal error estimates are given in parenthesis.}
\scriptsize
\renewcommand{\arraystretch}{1.1}
\label{tablc}
\begin{tabular}{lccc}	
\hline
Parameter      		     & 002821-1453.3 & 012450-3241.4 & 024155+2507.8 \\
\hline                                            
$i$~${({^\circ})}$ 	     & 78.2(6)	&	81.8(6) &		80.8(3) \\
$q$		   	             & 0.173(7) &	0.377(8) & 	0.45(1) \\
$T_{1}$~(K)	   	     & 6540 & 		5186 & 		5746  \\
$T_{2}$~(K)	   	     & 6593(79) & 	5164(25) & 	5537(61) \\
$\Omega_{1} = \Omega_{2}$    & 	2.12(2) & 	2.60(2) & 	2.76(3) \\
$f$ (\%)		     &   	40  & 	13 &	 7 \\
$\frac{L_{1}}{L_{1}+L_{2}}$  &  0.82(1)       	& 0.71(1)	       & 0.72(2) \\
$\overline{r_{1}}$	     &  0.549(9)       	& 0.472(7)             & 0.44(1)\\
$\overline{r_{2}}$	     &  0.26(3)        	& 0.30(1)              & 0.31(2) \\
$T_{0}$~(HJD-2450000)	     &  2080.8634(6)	& 4740.6713(2)         & 4409.6264(6) \\
$P$~(d)			     &  0.4026640(1)	& 0.30897005(2)        & 0.40088802(5)	     \\
\cline{2-4}
                              & 050334-2521.9   & 051353-1701.2         & 063546-1928.6   \\
\cline{2-4}                                
$i$~${({^\circ})}$ 	      &  83.3(2)	& 70.3(7) 		&  85.4(1)	\\
$q$		   	      &  0.133(3)	& 0.52(4)		&  0.173(4)	\\
$T_{1}$~(K)	   	      &  6347  		& 5419			&  6229		\\
$T_{2}$~(K)	   	      &  5925(40)	& 5086(53)		&  6072(29)	\\
$\Omega_{1} = \Omega_{2}$     &  2.01(1)	& 2.81(8)		&  2.10(1)	\\
$f$ (\%)		      &  53  		& 34			&  58		\\
$\frac{L_{1}}{L_{1}+L_{2}}$   & 0.885(5)	& 0.71(3)		&  0.834(4)	\\
$\overline{r_{1}}$	      & 0.572(5)	& 0.46(3)		&  0.557(4)	\\
$\overline{r_{2}}$	      & 0.24(3)  	& 0.35(4)		&  0.27(2)	\\
$T_{0}$~(HJD-2450000)	      & 3630.8372(4)	& 1914.6192(5)		&  3705.7085(3) \\
$P$~(d)			      & 0.41406990(4)	& 0.34183617(8)		&  0.47551487(3) \\
\cline{2-4}

\hline\end{tabular}
\end{center}
\end{table*}

\section{Individual objects}

\subsection{ASAS~002821-1453.3}
According to our $\it{V}$ light curve solution this system contains two components with temperature difference of about 50~K. Its light curve presents a typical W~UMa type shape. ($\it{V-K_{2}}$) = 1.0293 and ($\it{V_{T}-K_{2}}$) = 1.1072  colors correspond to the effective temperatures of 6566 K and 6514 K, respectively. The fillout factor of the system is about 40\%.  

\subsection{ASAS~012450-3241.4}
The target is the coolest system among our selected binaries. Its ($\it{V-K_{2}}$) = 1.9160 and ($\it{V_{T}-K_{2}}$) = 2.0320 colors indicate a primary having $T$ = 5203 K and  $T$ = 5169 K, respectively, according to our calculations. The results of our analysis show that temperature difference of the binary components is very small ($\sim$ 22 K) and one can assume that the two companions have almost equal effective temperature values. Its fillout factor is about 13\%. 

\subsection{ASAS~24155+2507.8}
The system has the smallest fillout factor among our targets with the value of 7\%. The temperature difference of two components is about 200 K. ($\it{V-K_{2}}$) = 1.5122 and ($\it{V_{T}-K_{2}}$) = 1.5861 colors correspond to the temperatures  $T$=5750 K and  $T$=5746 K, respectively.  

\subsection{ASAS~050334-2521.9}
The target appeared in the catalog of \cite{get06} with the orbital period value of 0.414075 days. Results of our analysis indicate that the system has a mass--ratio value of 0.133 and it can be put into a small group of deep, low mass--ratio contact binaries following the criteria of $f\geqslant$~50$\%$ given by \cite{qia05}. Its ($\it{V-K_{2}}$) = 1.1676 and ($\it{V_{T}-K_{2}}$) = 1.1945 color values show that temperature of primary component is $T$=6330 K and  $T$ = 6364 K, respectively. There is a $\sim$ 400 K temperature difference between the components of the binary system.

\subsection{ASAS~51353+1701.2}
 ($\it{V-K_{2}}$) and ($\it{V_{T}-K_{2}}$) colors of this system are 1.7336 and 1.8333, and the temperatures corresponding these values are $T$=5433 K and  $T$ = 5405 K, respectively. It has the largest mass--ratio with its value of 0.52 among our selected targets. The temperature difference of two components and its fillout factor are about 350 K  and 34\%, respectively.

\subsection{ASAS~063546+1928.6}
The light curve of the system shows total eclipses and therefore has two flat bottomed minimums. The system's inclination, 85$^\circ$.4, is the highest among our targets. The star can be put into the group of deep, low mass--ratio contact binaries \citep{qia05} owing to its degree of contact and mass--ratio values ($f$=58$\%$, $q$=0.173). ($\it{V-K_{2}}$) = 1.2140 and ($\it{V_{T}-K_{2}}$) = 1.2846 colors correspond to the temperatures  $T$=6247 K and  $T$=6212 K, respectively, and the temperature difference of two components is about 150 K.

\section{Conclusions}
We presented the first light curve solutions of six contact binaries selected from ASAS database. The initial parameters for temperatures during the analyses were chosen using the $V-K$ colors of the targets. The light parameters and the agreement between observations and analysis were also presented. The absolute parameters were determined by estimating the masses of the primary components using the correlation given by \cite{cox00}. These parameters are listed in Table~\ref{tababs}. The calculation of the degree of contact ($f$) for all systems indicates that our targets cover a large interval of this parameter (7-58$\%$).
\begin{table*}
\begin{center}
\tiny
\caption{Estimated absolute parameters of the systems.}
\renewcommand{\arraystretch}{1.2}
\label{tababs}
\begin{tabular}{lcccccccc}
\hline
ASAS Number    &M$_1$ (M$_{\odot}$)  &  M$_2$ (M$_{\odot}$) & R$_1$ (R$_{\odot}$)  &  R$_2$ (R$_{\odot}$)& $T_{1}$~(K) & $T_{2}$~(K) & L$_1$ (L$_{\odot}$) & L$_2$ (L$_{\odot}$)\\
\hline
002821-1453.3       &1.33          &0.230(9)        &1.49(2)    &0.60(2)        &6540 &6593(79) & 3.7(1)    & 0.70(4)          \\
012450-3241.4	    &0.79	   &0.298(6)        &0.96(1)	&0.58(3)	&5186 &5164(25) & 0.59(2)   & 0.21(3)		\\
024155+2507.8	    &0.97	   &0.430(1)	    &1.18(3)	&0.77(7)        &5746 &5537(61) & 1.37(7)   & 0.50(9)	       \\
050334-2521.9       &1.26	   &0.168(4)	    &1.54(1)	&0.60(1)	&6347 &5925(40) & 3.45(6)   & 0.40(2)         \\
051353-1701.2       &0.89	   &0.460(4)	    &1.07(7)	&0.80(1)	&5419 &5086(53)	& 0.90(1)   & 0.30(1)          \\
063546+1928.6       &1.19	   &0.206(5)	    &1.63(1)	&0.70(1)	&6229 &6079(29) & 3.60(6)   & 0.60(2)         \\
\hline
\end{tabular}
\end{center}
\end{table*}

The contact binary stars were divided into two subclasses, A-- and W--types, by \cite{bin70}. The author mentioned that A--type contacts have the hotter components which are occulted during the primary minimum while W--types are the systems where the hotter companions are the less massive ones. According to this criteria, all of our targets belong to A--type. 

\cite{luc68} purposed that contact binary systems have two components sharing common envelope whose temperature is nearly constant through the surface. Thermal relaxation oscillations are the cyclic mass transfer phases that are suggested for the evolution of the contact binaries \citep{fla76, luc76} although there are studies contrary to this idea. \cite{pac06} mentioned that the controversy probably rises from the binary systems having the period between 0.3 $<P<$ 1.2 days and are not in thermal contact. A-- and W--type subclasses were thought to have different evolutionary properties for a long time. \cite{mac96}, on the other hand, purposed three subclasses for contact binaries. According to their classification one group has hot components while the other groups, old and young late types, are characteristic W~UMa systems. Later, \cite{gaz08} advised that intermediate mass W--type systems evolve to high mass ratio A--type systems. They also concluded that the old and young late type contact binaries preserves their configuration in their whole phases of evolution. \cite{ste12} also suggested a scenario for cool contact binaries: the detached binaries having orbital period smaller than 2 days evolve to a system transferring mass from the more massive component to less massive one and then to a contact binary. The evolution finalizes with the merging process which produces a rapidly rotating single star.
\begin{figure*}
\begin{center}
\begin{tabular}{|c|c|}
\hline
\includegraphics[scale=0.3]{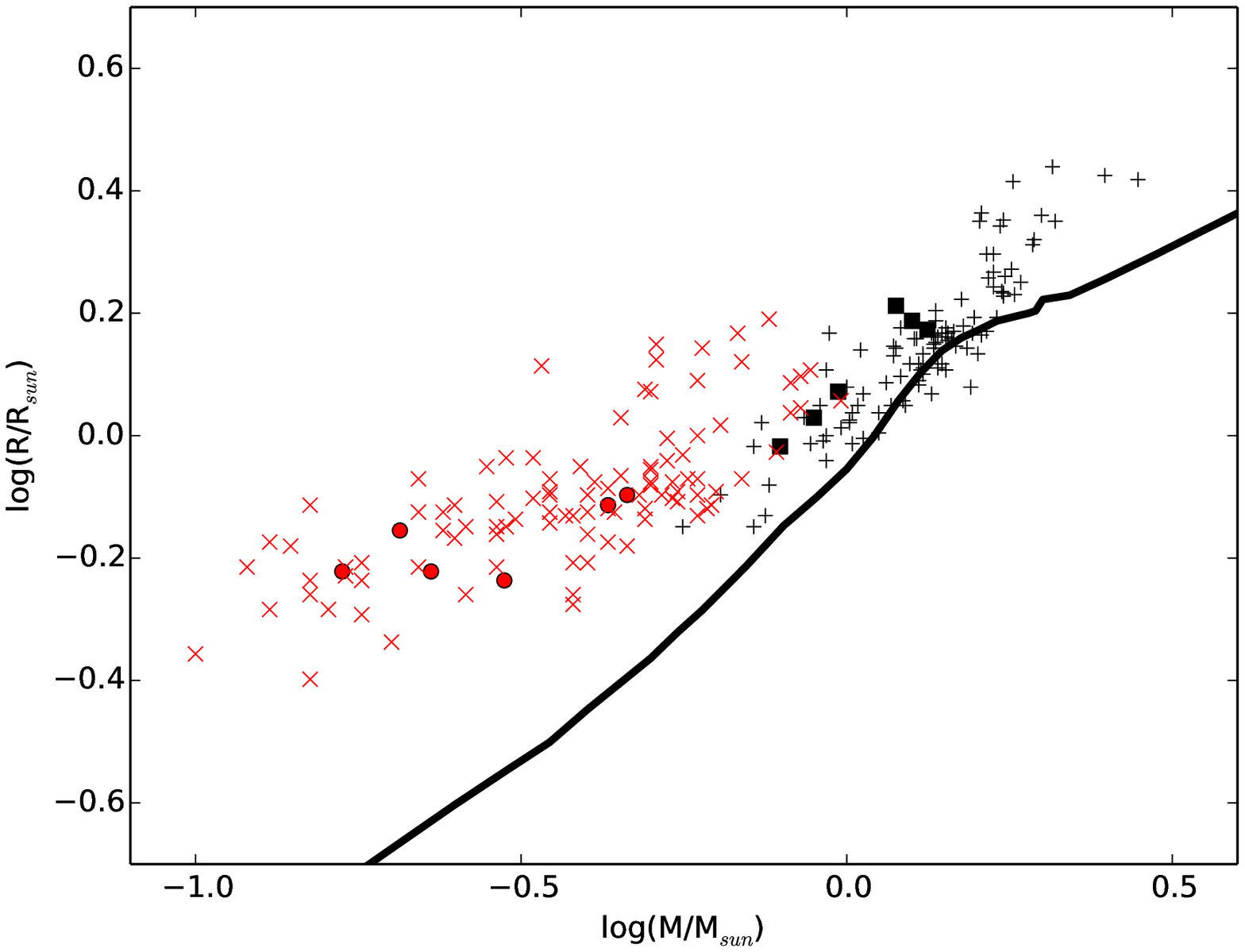} & \includegraphics[scale=0.3]{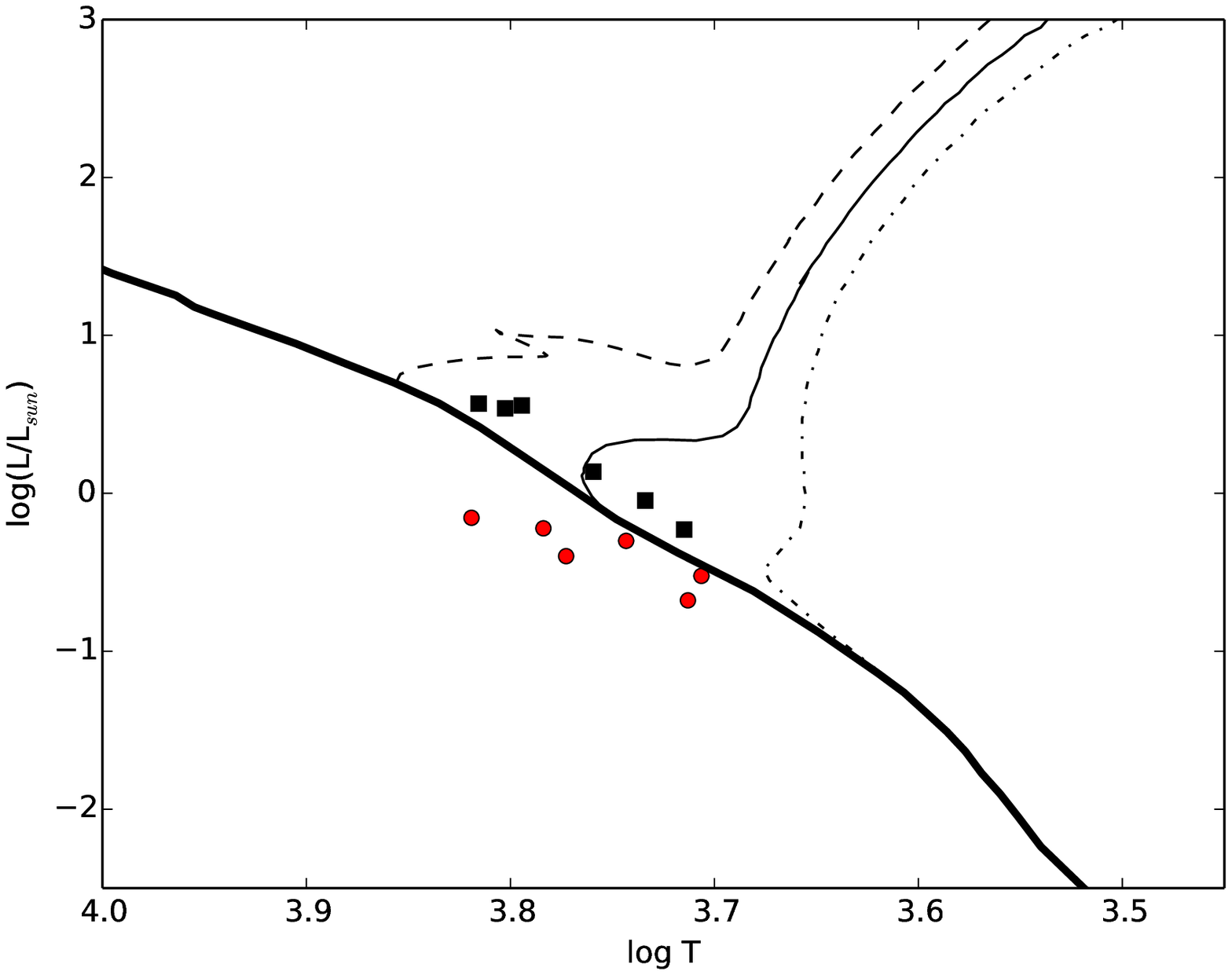} \\
\hline
\end{tabular}
\caption{The locations of the targets on mass--radius plane (left) and Hertzsprung--Russell diagram (right). The squares (black in coloured edition) indicate the primary components while the filled circles (red in coloured edition) denote the secondaries of our targets. The plus (black in coloured edition) and cross signs (red in coloured edition) on mass--radius plane correspond to primary and secondary components of contact binaries collected from \cite{yil13}. The thick solid lines indicate the zero age main sequence (ZAMS). The dashed, thin solid and dot--dashed lines show the evolutionary tracks of the stars having the initial mass of 1.5M$_{\odot}$, 1M$_{\odot}$ and 0.6M$_{\odot}$. The data for ZAMS and the evolutionary tracks are taken from \cite{gir00}.}
\label{figmrhr}
\end{center}
\end{figure*}
We compared the results of the analyses for our targets with the well known contact binaries. To make this comparison, we located the systems on mass--radius plane together with 100 contact binaries whose physical parameters were given by \cite{yil13}. Fig.~\ref{figmrhr} shows the locations of our targets together with the other contact binaries. The primary and secondary components are gathered in two separate regions in an accordance with other contact binaries. Therefore, our targets are in good agreement with other contact binaries in the plane. The positions of the components of the targets on the Hertzsprung--Russell diagram are also plotted in Fig.~\ref{figmrhr}. The secondaries are generally seen slightly under the main sequence while the primaries are on it as hypothesize. The evolutionary status of our targets shows that primary components of some of our targets seem more evolved than other ones which can be seen by the help of the evolutionary tracks drawn for the stars having the initial mass of 1.5M$_{\odot}$, 1M$_{\odot}$ and 0.6M$_{\odot}$ on the Hertzsprung--Russell diagram.
\section*{Acknowledgements}
This research has made use of the SIMBAD (operated at CDS, Strasbourg, France) database. Besides the ASAS catalogue, 2MASS and Tycho catalogues also helped us to determine the colors of our targets.


\begin{thebibliography}{}

\bibitem[\protect\citeauthoryear{Bernhard}{2004}]{ber04} Bernhard K., 2004, BAVSR, 53, 108 
\bibitem[\protect\citeauthoryear{Binnendijk}{1970}]{bin70} Binnendijk L., 1970, VA, 12, 217
\bibitem[\protect\citeauthoryear{Cox}{2000}]{cox00} Cox A.~N., 2000, asqu.book
\bibitem[\protect\citeauthoryear{de Geus, Lub, \& van de Grift}{1990}]{geu90} de Geus E.~J., Lub J., van de Grift E., 1990, A\&AS, 85, 915
\bibitem[\protect\citeauthoryear{Flannery}{1976}]{fla76} Flannery B.~P., 1976, ApJ, 205, 217
\bibitem[\protect\citeauthoryear{Flower}{1996}]{flo96} Flower P.~J., 1996, ApJ, 469, 355
\bibitem[\protect\citeauthoryear{Gazeas \& St{\c e}pie{\'n}}{2008}]{gaz08} Gazeas K., St{\c e}pie{\'n} K., 2008, MNRAS, 390, 1577
\bibitem[\protect\citeauthoryear{Gettel, Geske \& McKay}{2006}]{get06} Gettel S.~J., Geske M.~T., McKay T.~A., 2006, AJ, 131, 621
\bibitem[\protect\citeauthoryear{Girardi et al.}{2000}]{gir00} Girardi L., Bressan A., Bertelli G., Chiosi C., 2000, A\&AS, 141, 371 
\bibitem[\protect\citeauthoryear{Hoffman, Harrison, \& McNamara}{2009}]{hof09} Hoffman D.~I., Harrison T.~E., McNamara B.~J., 2009, AJ, 138, 466 
\bibitem[\protect\citeauthoryear{Kazarovets et al.}{2011}]{kaz11} Kazarovets E.~V., Samus N.~N., Durlevich 
O.~V., Kireeva N.~N., Pastukhova E.~N., 2011, IBVS, 5969, 1 
\bibitem[\protect\citeauthoryear{Castelli 
\& Kurucz}{2003}]{kurucz03} Castelli F., Kurucz R.~L., 2003, IAUS, 210, 20P 
\bibitem[\protect\citeauthoryear{Koch et al.}{1979}]{koc79} Koch R.~H., Wood F.~B., Florkowski D.~R., Oliver J.~P., 1979, IBVS, 1708, 1
\bibitem[\protect\citeauthoryear{Koz{\l}owski, Konacki, \& Sybilski}{2011}]{koz11} Koz{\l}owski S.~K., Konacki M., Sybilski P., 2011, MNRAS, 416, 2020
\bibitem[\protect\citeauthoryear{Landolt}{1992}]{Lan92} Landolt, A. U., 1992, AJ, 104, 340
\bibitem[\protect\citeauthoryear{Lopez \& Girard}{1990}]{lop90} Lopez C.~E., Girard T.~M., 1990, PASP, 102, 1018
\bibitem[\protect\citeauthoryear{Lucy}{1968}]{luc68} Lucy L.~B., 1968, ApJ, 151, 1123 
\bibitem[\protect\citeauthoryear{Lucy}{1976}]{luc76} Lucy L.~B., 1976, ApJ, 205, 208
\bibitem[\protect\citeauthoryear{Lucy  \& Wilson}{1979}]{luc79} Lucy L.~B., Wilson R.~E., 1979, ApJ, 231, 502 
\bibitem[\protect\citeauthoryear{Maceroni \& van't Veer}{1996}]{mac96} Maceroni C., van't Veer F., 1996, A\&A, 311, 523
\bibitem[\protect\citeauthoryear{Malkov et al.}{2006}]{mal06} Malkov O.~Y., Oblak E., Snegireva E.~A., Torra J., 2006, A\&A, 446, 785 
\bibitem[\protect\citeauthoryear{Marraco \& Orsatti}{1982}]{mar82} Marraco L.~G., Orsatti A.~M., 1982, RMxAA, 5, 183
\bibitem[\protect\citeauthoryear{Muzzio \& Orsatti}{1977}]{muz77} Muzzio J.~C., Orsatti A.~M., 1977, AJ, 82, 345
\bibitem[\protect\citeauthoryear{Paczy{\'n}ski et al.}{2006}]{pac06} Paczy{\'n}ski B., Szczygie{\l} D.~M., 
Pilecki B., Pojma{\'n}ski G., 2006, MNRAS, 368, 1311
\bibitem[\protect\citeauthoryear{Perryman et al.}{1997}]{Perry97} Perryman, M. A. C., Perryman, M. A. C.,  
Kovalevsky, J., Hoeg, E., Bastian, U. and 15 coauthors, 1997, A\&A, 323L, 49P
\bibitem[\protect\citeauthoryear{Philip \& Stock}{1972}]{phi72} Philip A.~G.~D., Stock J., 1972, BOTT, 6, 201 
\bibitem[\protect\citeauthoryear{Pojmanski}{1997}]{poj97} Pojmanski G., 1997, AcA, 47, 467 
\bibitem[\protect\citeauthoryear{Pojmanski}{2000}]{poj00} Pojmanski G., 2000, AcA, 50, 177 
\bibitem[\protect\citeauthoryear{Pols et al.}{1995}]{pol95} Pols O.~R., Tout C.~A., Eggleton P.~P., Han Z., 1995, MNRAS, 274, 964
\bibitem[\protect\citeauthoryear{Popova \& Kraicheva}{1984}]{pop84} Popova M., Kraicheva Z., 1984, AISAO, 18, 6
\bibitem[\protect\citeauthoryear{Pr{\v s}a \& Zwitter}{2005}]{prs05} Pr{\v s}a A., Zwitter T., 2005, ApJ, 628, 426
\bibitem[\protect\citeauthoryear{Qian et al.}{2005}]{qia05} Qian S.-B., Yang Y.-G., Soonthornthum B., Zhu L.-Y., He J.-J., Yuan J.-Z., 2005, AJ, 130, 224
\bibitem[\protect\citeauthoryear{Ram{\'i}rez \& Mel{\'e}ndez}{2005}]{Ram05} Ram{\'i}rez, I., Mel{\'e}ndez, J., 2005, ApJ, 626, 465
\bibitem[\protect\citeauthoryear{Ruci{\'n}ski}{1969}]{ruc69} Ruci{\'n}ski S.~M., 1969, AcA, 19, 245
\bibitem[\protect\citeauthoryear{St{\c e}pie{\'n} \& Gazeas}{2012}]{ste12} St{\c e}pie{\'n} K., Gazeas K., 2012, AcA, 62, 153
\bibitem[\protect\citeauthoryear{van Hamme}{1993}]{ham93} van Hamme W., 1993, AJ, 106, 2096 
\bibitem[\protect\citeauthoryear{Wenger et al.}{2000}]{wen00} Wenger M., Ochsenbein, F., Egret, D., Dubois, P.,  Bonnarel, F. and 6 coauthors, 2000, A\&AS, 143, 9
\bibitem[\protect\citeauthoryear{Wilson \& Devinney}{1971}]{wil71} Wilson R.~E., Devinney E.~J., 1971, ApJ, 166, 605
\bibitem[\protect\citeauthoryear{Y{\i}ld{\i}z \& Do{\u g}an}{2013}]{yil13} Y{\i}ld{\i}z M., Do{\u g}an T., 2013, MNRAS, 430, 2029 

\end{thebibliography}
\end{document}